\documentclass[twocolumn,prd,floatfix,preprintnumbers,a4paper,nofootinbib,superscriptaddress]{revtex4}

\usepackage{epsfig}
\usepackage{graphics}
\usepackage{graphicx}
\usepackage{amsmath,amssymb}
\usepackage{amsfonts}
\usepackage{bm}
\usepackage[usenames,dvipsnames]{color}
\usepackage{amssymb}
\usepackage{psfrag}
\usepackage{times}
\usepackage[varg]{txfonts}
\usepackage[colorlinks, pdfborder={0 0 0}]{hyperref}
\definecolor{LinkColor}{rgb}{0.75, 0, 0}
\definecolor{CiteColor}{rgb}{0, 0.5, 0.5}
\definecolor{UrlColor}{rgb}{0, 0, 0.75}
\hypersetup{linkcolor=LinkColor}
\hypersetup{citecolor=CiteColor}
\hypersetup{urlcolor=UrlColor}
\usepackage[utf8]{inputenc}
\usepackage{ulem}
\newcommand{\dd}{\mathrm{d}}
\newcommand{\Pimr}{P_{\textsc{imr}}}
\newcommand{\Pin}{P_{\textsc{i}}}
\newcommand{\Prd}{P_{\textsc{mr}}}
\newcommand{\Pdel}{P_{\Delta}}
\newcommand{\hgr}{h_\textsc{gr}}
\newcommand{\thgr}{\tilde{h}_\textsc{gr}}
\newcommand{\bS}{\boldsymbol{S}}
\newcommand{\lalinf}{\textsc{LALInference}}
\newcommand{\lalinferencenest}{\textsc{LALInferenceNest}}
\newcommand{\lal}{\textsc{LAL}}

\begin{document}

\title{Testing general relativity using golden black-hole binaries}

\author{Abhirup~Ghosh}
\author{Archisman~Ghosh}
\author{Nathan~K.~Johnson-McDaniel}
\author{Chandra~Kant~Mishra}
\author{Parameswaran~Ajith}
\affiliation{International Centre for Theoretical Sciences, Tata Institute of Fundamental Research, Bangalore 560012, India}
\author{Walter~Del~Pozzo}
\affiliation{School of Physics and Astronomy, University of Birmingham, Edgbaston, Birmingham, B15 2TT, United Kingdom}
\author{David~A.~Nichols}
\affiliation{Cornell Center for Astrophysics and Planetary Science (CCAPS),
Cornell University, Ithaca, New York 14853, USA}
\author{Yanbei~Chen}
\affiliation{Theoretical Astrophysics, California Institute of Technology, Pasadena, CA 91125, USA}
\author{Alex~B.~Nielsen}
\affiliation{Max-Planck-Institut f\"ur Gravitationsphysik, Albert-Einstein-Institut, Callinstr.38, 30167 Hannover, Germany}
\author{Christopher~P.~L.~Berry}
\affiliation{School of Physics and Astronomy, University of Birmingham, Edgbaston, Birmingham, B15 2TT, United Kingdom}
\author{Lionel~London}
\affiliation{School of Physics and Astronomy, Cardiff University, The Parade, Cardiff, CF24 3AA, United Kingdom}


\begin{abstract}
The coalescences of stellar-mass black-hole binaries through their inspiral, merger, and ringdown are among the most promising sources for ground-based gravitational-wave (GW) detectors. If a GW signal is observed with sufficient signal-to-noise ratio, the masses and spins of the black holes can be estimated from just the inspiral part of the signal. Using these estimates of the initial parameters of the binary, the mass and spin of the final black hole can be uniquely predicted making use of general-relativistic numerical simulations. In addition, the mass and spin of the final black hole can be independently estimated from the merger--ringdown part of the signal. If the binary black hole dynamics is correctly described by general relativity (GR), these independent estimates have to be consistent with each other. We present a Bayesian implementation of such a test of general relativity, which allows us to combine the constraints from multiple observations. Using kludge modified GR waveforms, we demonstrate that this test can detect sufficiently large deviations from GR, and outline the expected constraints from upcoming GW observations using the second-generation of ground-based GW detectors. 
\end{abstract}

\preprint{LIGO-P1500185-v11}
\date{\today}

\maketitle
\section{Introduction} 
The coalescence of black-hole binaries, driven by the emission of gravitational radiation, is perhaps the most luminous phenomenon occurring in the Universe after Big Bang. During the final stages of the coalescence, up to $\sim 10\%$ of the mass-energy of the binary is radiated as gravitational waves (GWs) over the last few orbits of the inspiral and merger (see, e.g.,~\cite{Centrella:2010mx} for a review). This will allow the second-generation ground-based GW observatories~\cite{AdvLIGO,acernese2015advanced,somiya2012detector,LIGOIndiaProposal:2011,Aasi:2013wya} to detect such phenomena up to distances of several gigaparsecs~\cite{Ajith:2007kx}, making binary black hole coalescences some of the most promising sources of GWs for these observatories. Expected detection rates for second-generation detectors vary from a handful to several thousands per year, as predicted by population synthesis models~\cite{Dominik:2014yma,PhysRevLett.115.051101}.  Third-generation detectors~\cite{punturo2010einstein,miao2014quantum,dwyer2015gravitational} are expected to extend the range even further.

GW observations of binary black holes will enable us to test general relativity (GR) in a regime that is currently inaccessible by astronomical observations and laboratory tests. Apart from putting bounds on parameters of specific alternative theories, proposed tests include constraining parametrized deviations from post-Newtonian gravity, tests of the no-hair theorem by observing multiple quasi-normal modes or by constraining deviations from the expected multipolar structure of black holes, etc.\ (see, e.g.,~\cite{Berti:2015itd,Yunes:2013dva} for reviews). Here we present a test of GR based on GW observations of ``golden'' black-hole binaries~\cite{Hughes:2004vw,Nakano:2015uja} -- binaries with total mass $\sim 50 M_\odot$--$200 M_\odot$, so that the signals observed by ground-based GW observatories cover the inspiral, merger and ringdown (IMR) phases of the coalescence. During the inspiral, the two black holes spiral-in under gravitational radiation reaction, and eventually merge to form a common horizon. In the ringdown stage, the newly formed horizon settles into a Kerr black hole with the emission dominated by a spectrum of quasi-normal modes. According to the no-hair theorem, the final black hole is fully characterized by its mass and spin angular momentum. 

The idea of the proposed test is that, if a GW signal is observed with sufficient signal-to-noise ratio (SNR), the masses and spins of the black holes can be estimated just from the inspiral part of the signal. Given the estimates of the initial parameters of the binary, the mass and spin of the final black hole can be uniquely predicted making use of fits to numerical-relativity (NR) simulations. In the same way, the mass and spin of the final black hole can be independently estimated from the merger--ringdown portion of the signal.\footnote{The original test proposed in \cite{Hughes:2004vw} in the context of LISA makes use of only the inspiral and ringdown parts. In the case of second-generation ground-based detectors, the ringdown SNR is unlikely to be large for most events. Luckily, recent advances in NR have allowed us to model the merger accurately. Hence, our implementation makes use of the merger part as well. However, it is possible to restrict our test solely to the inspiral and ringdown parts by appropriate choice of the cutoff frequencies defined in Eq.~\eqref{eq:likelihood}.} If the binary black hole dynamics is correctly described by GR, these independent estimates have to be consistent with each other. The consistency of the parameters estimated from the highly relativistic post-inspiral regime with those inferred from the weakly relativistic inspiral regime is a nontrivial test of the ability of GR in modeling this complex phenomenon.  

\section{Formulation of the test} 
The set of parameters $\boldsymbol{\lambda}$ of the binary, such as the masses ($m_1, m_2$) and spin angular momenta ($\bS_1,\bS_2$) of the black holes, are imprinted on the gravitational waveform. Given data $d(t)$ containing an observed GW signal, and assuming the GR model  $\hgr$, the posterior distribution $P(\boldsymbol{\lambda}|d,\hgr)$ of these parameters can be estimated making use of Bayes' theorem
\begin{equation}
P(\boldsymbol{\lambda}|d, \hgr) = N^{-1} \, p(\boldsymbol{\lambda}) \, \mathcal{L}(d|\hgr, \boldsymbol{\lambda}),
\end{equation}
where $p(\boldsymbol{\lambda})$ is the prior distribution of $\boldsymbol{\lambda}$, $N$ is a normalization constant (called the evidence) and $\mathcal{L}(d|\hgr, \boldsymbol{\lambda})$ is the likelihood of observing the data $d$ given the signal model $\hgr$ and the set of parameters $\boldsymbol{\lambda}$,
\begin{equation}
\mathcal{L} = \exp \left[-\int_{f_\mathrm{low}}^{f_\mathrm{up}} \frac{|\tilde{d}(f)-\thgr(f,\boldsymbol{\lambda})|^2}{S(f)} \dd f \right].
\label{eq:likelihood}
\end{equation}
Above, $\tilde{d}(f)$ is the Fourier transform of the data, $\thgr(f,\boldsymbol{\lambda})$ is the frequency-domain signal waveform corresponding to the set of parameters $\boldsymbol{\lambda}$, and $S(f)$ is the power spectral density of the detector noise, while $f_\mathrm{low}$ and $f_\mathrm{up}$ are the lower and upper cutoff frequencies used in the calculation. The sampling of the likelihood function $\mathcal{L}(d|\hgr, \boldsymbol{\lambda})$ over the (typically large dimensional) parameter space often makes use of stochastic sampling methods such as Markov-chain Monte-Carlo or nested sampling~\cite{LALInference}. 

First, we estimate the joint posterior probability $\Pimr(m_1, m_2, \bS_1, \bS_2)$ (marginalized over all other parameters of the binary) from the complete observed IMR signal.\footnote{From here onwards, we drop the explicit reference to the data $d$ and the GR model $\hgr$ in the posteriors, for simplicity.} This allows us to infer the posterior $\Pimr(M_f,\chi_f)$ on the mass $M_f$ and dimensionless spin $\chi_f := |\bS_f|/M^2_f$ of the final black hole, using fitting formulas (e.g.,~\cite{Healy:2014yta}) calibrated to NR simulations 
\begin{equation}
M_f = M_f(m_1, m_2, \bS_1, \bS_2), ~~~~~ \chi_f = \chi_f(m_1, m_2, \bS_1, \bS_2).
\label{eq:spin_fit_formula}
\end{equation}
We use these estimates of $M_f$ and  $\chi_f$ to split the signal into an inspiral part and a merger--ringdown part. In this paper, we define the inspiral [merger--ringdown] part as Fourier frequencies less [greater] than that of the innermost stable circular orbit (ISCO) of a Kerr black hole with mass and spin equal to that given by the median value of $\Pimr(M_f,\chi_f)$.\footnote{While we split the signal in the Fourier domain, we have checked that almost all the power below [above] our split frequency indeed comes from the early [late] portions of the waveform; the effect of the spectral leakage is negligible.} However, this choice is not unique; alternative ways of splitting the signal are possible, and reasonable alternatives do not have a significant effect on the test. 

\begin{figure}[tb]
\includegraphics[width=0.48\textwidth]{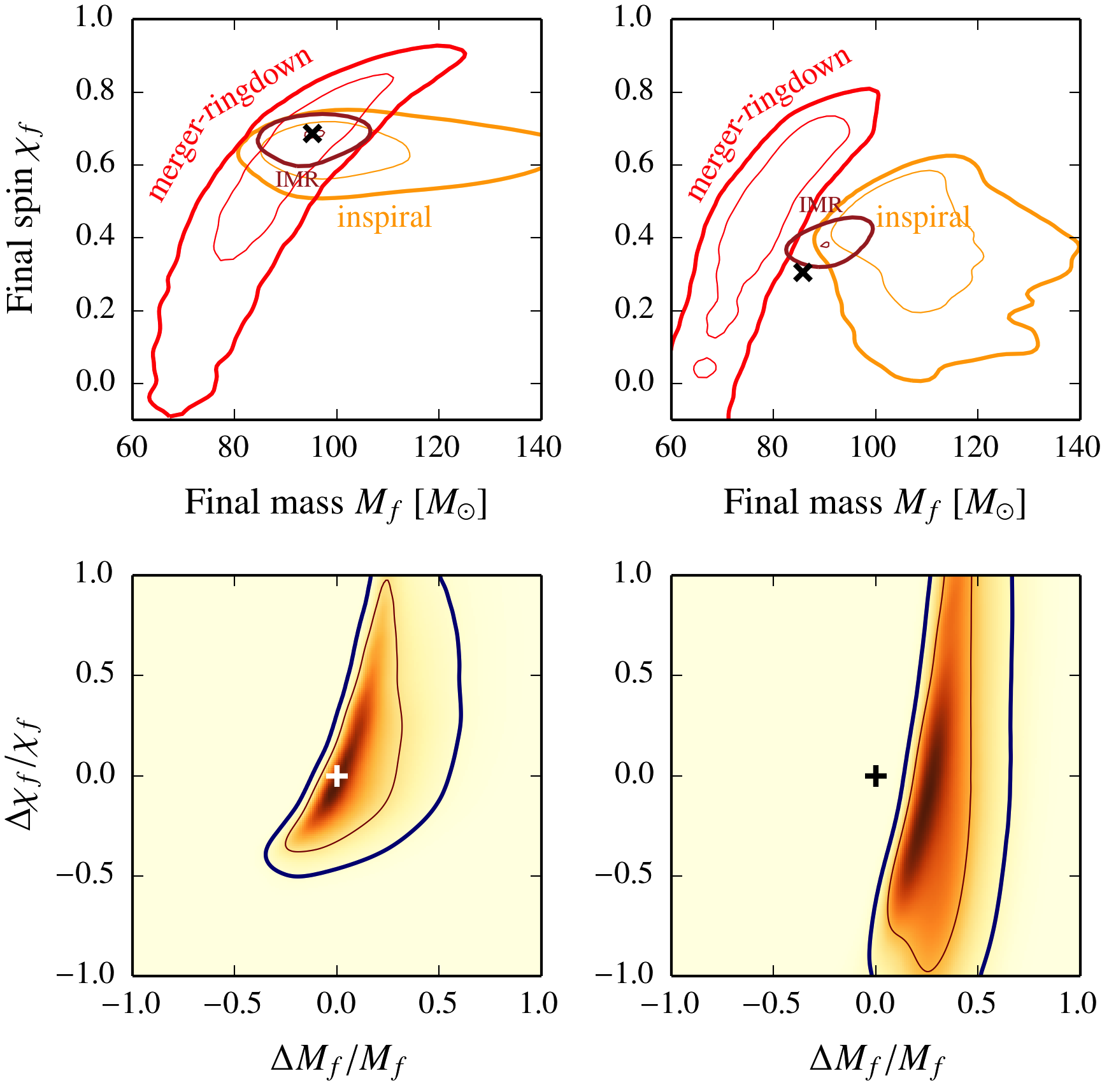}
\caption{\emph{Left panels:} The top left panel shows the $68\%$ and $95\%$ credible regions of the posterior distributions $\Pin(M_f,\chi_f)$ and $\Prd(M_f,\chi_f)$ of the mass and spin of the final black hole estimated from the inspiral and merger--ringdown parts of a simulated GR signal, respectively. Also shown is the posterior $\Pimr(M_f, \chi_f)$ estimated from the full IMR signal. The simulated GR signal is from a non-spinning black hole binary with $m_1 = m_2 =  50 M_\odot$, producing an optimal SNR of 25 in the Advanced LIGO Hanford--Livingston network. The corresponding value of the final mass and spin is indicated by a black cross. The bottom left panel shows the posterior $P(\epsilon, \xi)$ on the parameters $\epsilon := \Delta M_f/M_f$ and  $\xi := \Delta \chi_f/\chi_f$ that describe the deviation from GR, estimated from the same simulation. The GR value is marked by a ``+'' sign; the posterior is consistent with the GR value. \emph{Right panels:} Same as the left panels, except that here the injection corresponds to a modified GR signal with $\alpha_\mathrm{modGR} = 400$, with the location and orientation of the binary same as that of the left panels, thus producing an optimal SNR of $18.9$. The GR value is well outside the $95\%$ credible region. In this example, GR can be ruled out with confidence $\gg 99\%$.}
\label{fig:single_posterior}
\end{figure}

We can now independently estimate the posterior $\Pin(m_1, m_2, \bS_1, \bS_2)$ from the inspiral part of the signal and compute the corresponding posterior $\Pin(M_f,\chi_f)$ of the mass and spin of the final black hole using the fitting formula Eq.~\eqref{eq:spin_fit_formula}. We independently estimate the posterior $\Prd(M_f,\chi_f)$ from the merger--ringdown part of the signal. In the absence of any deviations from GR (or significant systematic errors), we expect the two posteriors $\Pin(M_f,\chi_f)$ and $\Prd(M_f,\chi_f)$ to overlap (see, e.g., the top left panel in Fig.~\ref{fig:single_posterior}).

\begin{figure*}[tb]
\includegraphics[width=\textwidth]{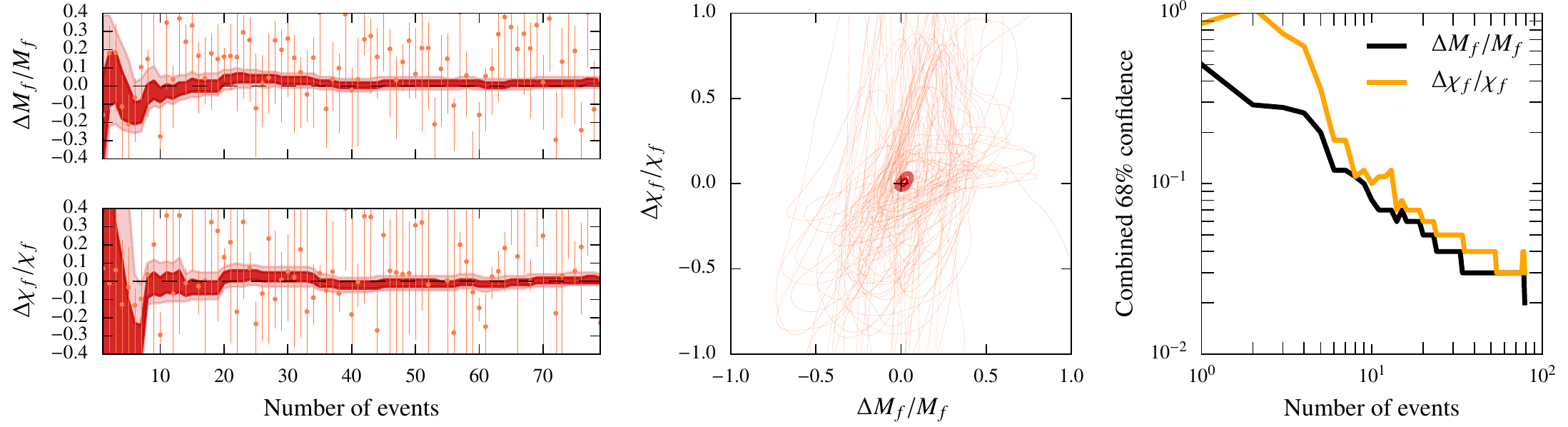}
\caption{\emph{Left panels:} Shaded regions show the $68\%$ and $95\%$ credible intervals on the combined posteriors on $\epsilon, \xi$ from multiple observations of GR signals plotted against the number of observations by Advanced LIGO. The GR value ($\epsilon = \xi = 0$) is indicated by horizontal dashed lines. The mean value of the posterior from each event is shown as an orange dot along with the corresponding $68\%$ credible interval. Posteriors on $\epsilon$ are marginalized over $\xi$, and vice versa. \emph{Middle panel:} The orange contours show the $68\%$ credible regions of the individual posteriors on the $\epsilon, \xi$ computed from the same events while the thick red contour shows the $68\%$ and $95\%$ credible regions on the combined posterior. \emph{Right panel:} The width of the $68\%$ credible region in the marginalized posteriors of $\Delta M_f/M_f$ and $\Delta \chi_f/\chi_f$ from multiple observations.}
\label{fig:conf_intervals_combined}
\end{figure*}

To constrain possible departures from GR, we define two parameters that describe departures from the GR prediction of the mass and spin of the final black hole 
\begin{equation}
\Delta M_f := M^\textsc{i}_f - M^\textsc{mr}_f, ~~~~~~ \Delta \chi_f := \chi^\textsc{i}_f - \chi^\textsc{mr}_f,
\end{equation}
whose posterior distribution can be computed as 
\begin{eqnarray}
\Pdel(\Delta M_f, \Delta \chi_f) & = & \iint \dd M_f \,\dd\chi_f \, \Pin(M_f,\chi_f) \, \times \, \nonumber \\ 
 &&\qquad \Prd(M_f-\Delta M_f,\chi_f-\Delta \chi_f).
\label{eq:P_delta}
\end{eqnarray}
In the absence of departures from GR, we expect $P(\Delta M_f,\Delta \chi_f)$ to be consistent with zero. We define two quantities $\epsilon := \Delta M_f/M_f$ and $\xi := \Delta \chi_f/\chi_f$ that describe the fractional differences in the two predictions of the mass and spin of the final black hole. The posteriors on these can be computed as 
\begin{equation}
P(\epsilon, \xi) = \iint \dd M_f \,\dd\chi_f \, \Pdel(\epsilon M_f,\xi \chi_f) \, \Pimr(M_f,\chi_f) \, M_f \, \chi_f . 
\label{eq:P_epsilon_xi}
\end{equation}
Here, the posterior $\Pimr(M_f,\chi_f)$ denotes our best estimate of the mass and spin of the final black hole assuming GR, which is estimated from the full IMR waveform. An example of the posterior distribution $P(\epsilon,\xi)$ from a simulated GR signal is shown in the bottom left panel of Fig.~\ref{fig:single_posterior}. Finally, the posteriors $P(\epsilon,\xi)$ from multiple observations of binary black holes can be combined to construct a single posterior that can better constrain deviations from GR (see, e.g., Fig.~\ref{fig:conf_intervals_combined}).

\section{Implementation}
To compute the posterior distributions, we employ the \lalinf\ \cite{LALInference} stochastic samplers available in the LIGO Algorithm Library (\lal) \cite{urlLAL}. In particular we use the \lalinferencenest\ code~\cite{Veitch:2009hd}, which implements a nested sampling algorithm \cite{Skilling2004a} in the context of GW data analysis. As the GR signal model we employ the gravitational waveform family \textsc{SEOBNRv2\_ROM\_DoubleSpin}~\cite{Purrer:2015tud} which describes the inspiral, merger and ringdown waveform of black-hole binaries with non-precessing spins. This is a reduced-order model version~\cite{Purrer:2015tud} of the effective-one-body (EOB) waveform family~\cite{Taracchini:2013rva} calibrated to NR simulations. We use the fitting formulas proposed in \cite{Healy:2014yta} to compute the mass and spin of the final black hole from the initial masses and (non-precessing) spins.  

From the (simulated) data containing a GW signal, we compute the posterior distributions of $M_f$ and $\chi_f$ in three different ways: 
\begin{enumerate}
\item $\Pimr(M_f,\chi_f)$ is computed from the full data: we set $f_\mathrm{low} = f_0$ and $f_\mathrm{up} = f_\mathrm{Nyq}$ in Eq.~\eqref{eq:likelihood}, where $f_0$ is the low-frequency cutoff of the detector and $f_\mathrm{Nyq}$ is the Nyquist frequency of the data. From the median value of the posterior $\Pimr(M_f,\chi_f)$, we compute the frequency of the Kerr ISCO ($f_\mathrm{ISCO}$). This is used as the characteristic frequency to delineate the inspiral and merger--ringdown parts of the signal in our current analysis. 
\item $\Pin(M_f,\chi_f)$ is computed from the inspiral part of the data: we set $f_\mathrm{low} = f_0$ and $f_\mathrm{up} = f_\mathrm{ISCO}$ in Eq.~\eqref{eq:likelihood}. 
\item $\Prd(M_f,\chi_f)$ is computed from the merger--ringdown part of the data: we set $f_\mathrm{low} = f_\mathrm{ISCO}$ and $f_\mathrm{up} = f_\mathrm{Nyq}$ in Eq.~\eqref{eq:likelihood}. 
\end{enumerate}
All posteriors are computed by assuming a prior distribution that is uniform in $M_f$ and $\chi_f$. The posterior $\Pdel(\Delta M_f,\Delta \chi_f)$ is computed from Eq.~\eqref{eq:P_delta} using SciPy's \textsc{correlate2d} function and $P(\epsilon, \xi)$ is computed by numerically integrating Eq.~\eqref{eq:P_epsilon_xi}.

\section{GR simulations}
We have performed simulations where we inject simulated GW signals  modelling inspiral, merger and ringdown of binary black holes (based on GR, as modelled by \textsc{SEOBNRv2\_ROM\_DoubleSpin}) into colored Gaussian noise with the design power spectrum of the Advanced LIGO detectors in the high-power, zero-detuning configuration~\cite{aLIGOPSD}, with a low frequency cutoff $f_0 = 10$ Hz. Binaries had component masses (detector frame) uniformly distributed in the range $m_{1,2} = [10, 80]~M_\odot$ and non-precessing spins in the range $\chi_{1,2} = [-0.98, 0.98]$. Sources were distributed uniformly in the sky with isotropic orientations in such a way that the observed signals will have a network SNR of $\sim 25$. The estimated posterior $P(\epsilon, \xi)$ from a single simulated event is shown in the bottom left panel of Fig.~\ref{fig:single_posterior}. We also combine posteriors from multiple events; Figure~\ref{fig:conf_intervals_combined} shows the combined posteriors $P(\epsilon, \xi)$ as a function of the number of simulated events. The constraints on the deviation parameters $\{\epsilon, \xi\}$ become narrower when multiple events are combined. The width of the $68\%$ credible region could be as low as a few percent when $\sim 100$ observations are combined. This is within the reach of one year of Advanced LIGO observation, according to several population synthesis models~\cite{Dominik:2014yma,PhysRevLett.115.051101}. 

\section{Modified GR simulations}
We also test our analysis pipeline using simulated GW signals that show departures from GR. To obtain waveforms whose energy and angular momentum loss differs from that predicted by GR, we have chosen to make kludge waveforms based on a simple modification of EOB waveforms. Specifically, we take the IHES EOB waveform model described in~\cite{Damour:2012ky}, which is given as publicly available code at~\cite{IHES_EOB_url}, and modify the GW flux starting at second post-Newtonian ($2$PN) order by multiplying the six modes that first enter at $2$PN [viz., the $(\ell,m) = (3, \pm2)$, $(4,\pm4)$, and $(4,\pm2)$ modes] by a constant factor $\alpha_\text{modGR} = 400$.\footnote{The corresponding change in the PN phasing coefficients will depend on the mass ratio. For equal masses, the 2PN term in the frequency domain phase expression will be modified by a factor of $\sim -13$.} Such a $2$PN modification to the flux is unconstrained by measurements of the GW energy loss from the double pulsar J0737$-$3039~\cite{KramerEtAl:2006,YunesHughes2010}. We also multiply those modes of the waveform by a consistent factor $\alpha_\text{modGR}^{1/2}$. However, only the dominant $(2,\pm2)$ modes are used for simulating the observation. As in the original code, we use the maximum of the orbital frequency (calculated from the EOB Hamiltonian) to mark the termination of the inspiral (and the start of the matching to the quasi-normal modes to give the merger and ringdown). The eccentricity of our modified waveforms remains as small ($\lesssim 10^{-5}$) as for the unmodified waveforms. 

Since the final mass and spin in the original EOB waveform are set by a fit to NR results, for the modified waveform we replace this determination by demanding self-consistency of the radiated energy and angular momentum. That is, we choose the final mass and spin by minimizing the difference between the values we set for the final black hole and those obtained by energy and angular-momentum balance using the initial data and the radiated quantities calculated from the waveform (through $\ell = 7$). This treatment assumes that the standard GR expressions for the radiated energy and angular momentum remain valid for this modified gravity waveform, which is indeed the case for a large range of modified theories~\cite{Stein:2010pn}. We have not changed the quasi-normal mode spectrum of the final black hole, for simplicity. 

The right panels of Fig.~\ref{fig:single_posterior} show the estimated posteriors on the mass and spin of the final black hole from one modified GR simulation (equal-mass, non-spinning binary), for which the final mass and spin are $85.7M_\odot$ and $0.307$, compared to $95.2M_\odot$ and $0.687$ in the analogous GR case.\footnote{The GR waveform used in the left panels of Fig.~\ref{fig:single_posterior} was computed using the unmodified IHES EOB code, to allow a direct comparison with the modified GR result, though the differences between SEOBNRv2 and IHES EOB are very small for this equal-mass, non-spinning case.} We also show the posterior $P(\epsilon,\xi)$ on the parameters describing deviations from the GR predictions. It can be seen that the GR value (marked by a ``+'' sign) is well outside the $95\%$ credible region of the estimated posterior. In this example, GR can be ruled out with $\gg 99\%$ confidence. We have verified that this signal, having an optimal SNR of 18.7, produces a chi-square weighted SNR $\simeq 15$ when filtering with the best-fit GR waveform and would thus likely be detected by a standard detection pipeline~\cite{Usman:2015kfa}.

\section{Conclusions}
The test that we propose assumes the validity of GR and tests the null hypothesis by computing the posterior distribution for the parameters $(\epsilon, \xi)$ that quantify a deviation from the result in GR, where both parameters are identically zero. Multiple observations could be combined to produce better constraints on the deviation. We have seen that this test is able to detect deviations from GR that are not constrained by radio observations of the orbital decay of the double pulsar -- the tightest constraint available. The test is not based on a specific theory and, consequently, could work in any theory in which massive compact binaries inspiral, merge, and then ringdown. Conversely, if the data were inconsistent with the null hypothesis, then they would not be able to give any direct indication of which modified theory is responsible for the deviation from GR. We expect this test to complement other GW-based tests of GR, including those looking for specific modifications to GR and those looking for generic parametrized deviations, providing confidence in any statements of whether a given signal (or population of signals) is consistent with GR. 

Although we have used the ISCO frequency of the final Kerr black hole to delineate between inspiral and merger--ringdown in this paper, alternative ways of splitting the signal are possible. We have verified that the main results are robust against (reasonable) choices of cutoff frequencies. We have neglected the effect of spin precession and subdominant modes in this paper. However, they can be readily included in this method by incorporating these effects in our GR model $\hgr$ and also (in the case of precession) in the fitting formulas for the final mass and spin. Systematic errors due to waveform inaccuracies could be mitigated or quantified by using waveform models that are better calibrated to NR simulations as they become available. Methods for mitigating the systematic errors due to detector calibration errors have been independently developed which involve marginalizing the posterior distributions of the masses and spins over additional parameters that model calibration errors~\cite{Farr:2016xx:prep}. Studies pertaining to these aspects are to be reported in a forthcoming paper~\cite{Ghosh:2016xx:prep}.

The test introduced in this paper has already had its first application: This was one of the tests used to establish the consistency of LIGO's first gravitational wave detection with a binary black hole signal as predicted by GR~\cite{Abbott:2016blz,TheLIGOScientific:2016src}.

\acknowledgments
We thank J.~Veitch and A.~Nagar for assistance with the \textsc{LALInference} and IHES EOB codes, respectively. We also thank K.~G.~Arun, A.~Buonanno, N.~Christensen, B.~R.~Iyer, C.~Messenger, A.~Mukherjee, B.~S.~Sathyaprakash and C.~Van Den Broeck for useful discussions. Ar.~G., N.~K.~J.-M., and P.~A. acknowledge support from the AIRBUS Group Corporate Foundation through a chair in ``Mathematics of Complex Systems'' at ICTS. P.~A.'s research was, in addition, supported by a Ramanujan Fellowship from the Science and Engineering Research Board (SERB), India, the SERB FastTrack fellowship SR/FTP/PS-191/2012, and by the Max Planck Society and the Department of Science and Technology, India through a Max Planck Partner Group at ICTS. W.~D.~P. was partly supported by a Leverhulme Trust research project grant. Y.~C.'s research is supported by NSF Grant PHY-1404569, and D.N.'s by NSF grant PHY-1404105. C.~P.~L.~B. was supported by the Science and Technology Facilities Council. Computations were performed at the ICTS clusters Mowgli, Dogmatix, and Alice. 

\bibliographystyle{apsrev-nourl}
\bibliography{TestGR}

\end{document}